\documentstyle{lamuphys}
\makeatletter
\let\chapter\hid@chapter
\makeatother
\begin{document}
\pagenumbering{arabic}
\title{Mass Extinctions vs. Uniformitarianism in Biological Evolution}

\author{Per Bak and Maya Paczuski}

\institute{Department of Physics, Brookhaven National Laboratory,
Upton NY 11973 USA}

\maketitle

\begin{abstract}
It is usually believed that Darwin's theory leads to a smooth gradual
evolution, so that mass extinctions must be caused by external shocks.
However, it has recently been argued that mass extinctions arise from the
intrinsic dynamics of Darwinian evolution. Species become extinct when
swept by intermittent avalanches propagating through the global ecology.
These ideas are made concrete through studies of simple mathematical models
of coevolving species. The models exhibit self-organized criticality and
describe some general features of the extinction pattern in the fossil record.
\end{abstract}
\section{Introduction}

The theory of uniformitarianism, or gradualism, was formulated in the
last century by the geophysicist Charles Lyell (1830) in his tome,
{\it Principles of Geology}.  According to this theory,
all change is caused by processes that we
currently observe which have worked at the same rate at all times. For
instance, Lyell proposed that landscapes are formed by gradual
processes, rather than catastrophes like Noah's Flood, and the features
that we see today were made by slow persistent processes with time
as the ``great enabler''  that  eventually makes large changes.
Uniformitarianism is a ``linear'' theory where the amount of change
is proportional to the amount of time that has elapsed.

At first sight, Lyell's uniformitarian view is reasonable.
The laws of physics are generally expressed in terms of smooth,
continuous equations.  Since these laws should describe all observable
phenomena, it is natural to expect that the phenomena which we observe
should also vary in a smooth and gradual manner. The opposing
philosophy, catastrophism, claims that change occurs through
sudden cataclysmic events.  Since catastrophism smacks of creationism,
with no connection to the natural sciences as we know them, it has
been largely rejected by the scientific community.

Charles Darwin (1910) adapted Lyell's ideas of gradualism
in an uncompromising way.
According to his theory, evolution proceeds through random mutations
followed by selection of the fitter variants. This slow process takes
place at all times and all places at a steady rate.  Darwin took it
for granted that such a process would necessarily force evolution
to follow a smooth, gradual path. Consequently, Darwin denied the
existence of mass extinctions where a large fraction of species would
abruptly disappear.

\subsection{Avalanches and Punctuated Equilibrium}

However, we know that many natural phenomena evolve intermittently 
(Bak and Paczuski 1995; Paczuski, Maslov, and Bak 1996).
The dynamics may follow a step-like pattern with long, dormant
plateaus where little change takes place interrupted by sudden bursts,
or avalanches, of concentrated activity. The magnitudes of the
bursts may extend over a wide
range. Even though uniformitarianism, as opposed to
catastrophism, has historically dominated both geology and
paleontology, prototypical examples of intermittent behavior lie in
these two domains.

\subsubsection{Earthquakes:}

For instance, the crust of the earth accommodates large, devastating
 earthquakes in which hundreds of thousands of people are killed. Most
 of the time the crust of the earth appears to be stable. These
 periods of stasis are punctuated by earthquakes or avalanches taking
 place on a fault structure that stores information about the history
 of the system.

In fact, the size distribution of earthquakes follows a simple power law known
as the Gutenberg-Richter law (1956).  The power
law demonstrates that earthquakes have no typical scale; otherwise the
distribution would have a signature at that scale.  The smooth
variation from small to large events suggests that a common dynamical
mechanism is responsible for all earthquakes, regardless of their
size. Volcanic eruptions constitute another intermittent phenomenon in
geophysics. Solar flares, pulsar glitches, and the formation of
neutron stars are examples of intermittent behavior in
astrophysics. All these phenomena are examples where avalanches of activity
exhibit power law distributions similar to the
Gutenberg Richter law.  There is no way to accommodate the power law
distribution for earthquake sizes within the framework of
a linear theory such as uniformitarianism.

\subsubsection{A Gutenberg-Richter Law for Extinctions:}

One might, therefore, suspect that Darwin's use of
 uniformitarianism in a theory of evolution may also need to be
 reexamined.  In fact, about twenty years ago, Gould and Eldredge
 (1977) proposed that biological evolution takes place in terms of
 punctuations, where many species become extinct and new species
 emerge, interrupting periods of low activity or stasis. Figure 1
 shows the record of extinction events as recorded by J. J. Sepkoski
 (1993).  These extinction events in biology are analogous to earthquakes
in geology.  Note the spikes of extinction events spanning a range of
 magnitudes. The largest events are associated with the Cambrian
 explosion 500 million years ago, and the Permian extinction 200
 million years ago. Raup (1986) has plotted similar data as a histogram
 (figure 2) where each column shows the number of 4 million periods
 with a given extinction intensity.  The smooth variation from
 the smallest to the largest extinctions indicates a common
 mechanism.  Actually, punctuated equilibrium usually refers
 to the intermittent dynamics of a single species, where morphological
 change is concentrated in short time periods interrupted by long
 periods of stasis.

\subsection{External Shocks: ``Bad Luck''}

The extinctions of species appear to take place simultaneously across
families; they ``march to the same drummer''. This could be explained if
mass extinctions were caused by large, exogenous cataclysms, i. e. if
extinctions were due to ``bad luck'' rather than ``bad genes''.  For
example, in the most prominent theory,
\cite{alvarez} suggest that the Cretaceous extinction event where the
dinosaurs disappeared was caused by a meteor hitting the earth some 55
million years ago. Indeed, a large crater was observed
near the Yucatan peninsula in Mexico.
However, in order for an
exogenous event such as a meteor to wipe out an entire species, it
must have a global effect over the entire area that the species occupies;
otherwise the impact would be insufficient to cause extinction, except
for species with small local populations.
In addition, extinctions of
species take place all the time without an external cause.
Extinctions are taking place right now!  These extinction events are
obviously not caused by a meteor. Some are known to be intrinsic to
evolution, being caused by humans.

In his book {\it Bad Genes or Bad Luck}, Raup (1982) distinguishes
between bad luck, extinctions from external sources, and bad genes,
extinctions due to intrinsically poor fitness.  Whether or not
external shocks play an important role in evolution, it is important
to understand the dynamics of biological evolution in the absence of
these shocks.  

\subsection{Evolution of Isolated vs. Many Interacting Species}

In early theories of evolution, by Fisher (1932)
and others, evolution of a single species in isolation was
considered. Individuals within each species mutate, 
leading to a distribution of
fitnesses, and the fitter variants were selected. This leads to a
fitness which always increases smoothly {\it ad infinitum}.  Many
biologists appear content with this state of affairs, and rarely is
the need for a more comprehensive theory expressed. For instance,
Maynard Smith (1993), in his book {\it The Theory of Evolution}
notices with great satisfaction that nothing important has
changed in the 35 years intervening between his first and second
editions.

However, Fisher's picture of a
species evolving in isolation does not
appear to us to be able to explain any of the intricacy, diversity,
and complexity of real life.  This is because evolution is a cooperative
phenomenon. Species form ecologies where they interact with each other
in a global ecology with predator-prey relationships and food webs.
For example, humans depend on the oxygen emitted by trees and other
plants.  It is quite likely that the interaction of many species in a
global ecology plays a more important role in evolution than the
specific behavior of a single or a few species in isolation.

Our approach is to consider biology as a large, dynamical system with
very many degrees of freedom.  Interactive
 systems like biology may exhibit emergent
behavior which is not obvious from the study of typical local
interactions and detailed behaviors of two or three species.  Indeed,
the population dynamics of a few interacting species has been
described in terms of coupled differential equations, known as
Lotka-Volterra, or replicator equations. These equations may lead to
interesting, chaotic behavior, but of course not to mass extinction or
punctuated equilibrium.  Traditional evolutionary theory
 may be able to explain the behavior
of a few generations involving a few hundred rats, but it can not
explain evolution on the largest scale in which trillions of organisms
interact throughout billions of years.

\section{Self-Organized Criticality}

A few years ago, Bak, Tang, and Wiesenfeld (1987, 1988) suggested that
large dynamical systems may organize themselves into a
highly poised ``critical'' state where intermittent avalanches of all
sizes occur. The theory, self-organized criticality (SOC), is a nonlinear
theory for how change takes place in large systems.  It has become a
general paradigm  for intermittency and criticality in
Nature.  Evolution to the critical state is unavoidable, and occurs
irrespective of the interactions on the smallest scales which we
can readily observe.  A visual example is a pile of sand on a flat
table, onto which sand is gradually added. Starting from a flat
configuration, the sand evolves into a steep pile which emits
sandslides, or avalanches, of all sizes. This idea has been
successfully applied to a variety of geophysical and astrophysical
phenomena, in particular to earthquakes where it provides a natural
explanation for the Gutenberg-Richter law. It is now broadly accepted
that earthquakes are a self-organized critical phenomenon
\cite{newman1}.  One may think of SOC as the underlying theory for
catastrophism.

Can this nonlinear picture be applied to biological evolution?  Even
if we accept Darwin's mechanism for evolution, it is difficult to
extract its consequences for macroevolution. In contrast to the basic
laws of physics which are described by equations such as Newton's
equations, or Maxwell's equations, there are no ``Darwin's Equations''
to solve, as one of our editors, Henrik Flyvbjerg, once pointed out.
It may seem rather hopeless to try to mathematically describe
macroevolution without the fundamental microscopic equations.  On the
other hand, we know from our experience with many body collective
phenomena  that statistical properties of the system at
large scales may be largely independent of small scale details.  This
is called ``universality.''  The interactions are more important than
the details of the entities which make up the system.  Universality belies
the usual reductionist approach in the physical sciences where features
at large scales are explained in terms of models at successively smaller
scales with more and more details included.  Universality is a way to
throw out almost all of these details.

Thus, our studies are based
on abstract mathematical models. The models cannot be expected to reproduce 
any specific detail that may actually be observed in nature, such as humans 
or elephants. The confrontation between theory and reality must 
take place on a statistical level. This is not unusual in the natural 
sciences. Quantum mechanics and thermodynamics are inherently 
statistical phenomena. Chaotic systems are unpredictable, so 
comparison with experiment or observations must also be on the 
statistical level. Indeed, the Gutenberg-Richter law is a statistical law 
which can be explained in terms of grossly over-simplified
 SOC models for earthquakes.
One might hope to be able to do the same for biology.

We shall argue that
life may be a self-organized critical phenomenon. Periods 
of stasis where evolutionary activity is low
are interrupted by coevolutionary avalanches where species become
extinct and new species emerge.  Good genes 
during periods of stasis are no guarantee for survival. Extinctions 
may take place not only due to ``bad luck'' from external effects, like
meteors, but 
also due to bad luck from freak evolutionary incidents endogenous to 
the ecology.  Biological evolution operating in the critical state can 
explain a variety of empirical observations, including the lifetime 
distribution of species and the occurrence of mass extinctions.

\section{Co-evolutionary Avalanches}

Stuart Kauffman of the Santa Fe Institute was among the first to
suggest that life might be a self-organized critical phenomena where
evolution takes place in terms of co-evolutionary avalanches of all
sizes.  Together with Sonke Johnsen \cite{kj}
 he studied complex models of very many species forming
an interactive ecology, the NKC-models. In these models, each species
evolves in a rough fitness landscape, with many local peaks, employing
a picture invented by Sewall Wright 50 years ago \cite{wright} (figure
3) in his seminal work, {\it The Shifting Balance Theory}.
Populations are modified by means of mutation and differential
selection towards higher fitness. Random mutations allow individuals
to cross barriers, formed by troughs of lower
 fitness and move to other maxima.  Then
they might initiate a population at or near the new maximum.

Each species can be thought of as a group of individuals in the
vicinity of some fitness peak, and may be represented by the genetic
code of one of those individuals. In Kauffman's models, the genetic
code is represented by a string of N bits or genes
(0011011....11101000). Each configuration has a fitness associated
with it, which can be calculated from an algorithm, the
NK-algorithm. The contribution to the fitness from each gene or
``trait'' depends on the state of K other genes.  The fitness
depends on the coupling between genes.  The NK models are generalized
spin glass models, invented by physicists to describe metastability and
frozen behavior of magnetic systems with random interactions.

The elementary single step is what could be called a ``mutation
 of a species''. Despite the fact that this notation may raise a red flag
among biologists, it will be used throughout this chapter.
In evolution, this step
is made by random mutations of individuals followed by selection 
of the fitter variant, and subsequent transformation of the entire 
population to that variant. 
  The landscape is traced out as the bits are varied. By randomly 
mutating one bit at the time, and accepting the mutation only if it leads 
to a higher fitness, the species will eventually reach a local peak 
from which it can not improve further from single mutations. Of 
course, by making many coordinated mutations the species can  
transform into something even more fit, but this is very unlikely. 
A species can not spontaneously develop wings and fly.

However, the fitness landscape of each species is not rigid; it is a
rubber landscape modified when other species in the ecology change
their properties.  For instance, the prey of some predator may grow
thicker skin (or become extinct), so that the fitness of the predator
is reduced.  Within the landscape picture, this corresponds to a
deformation where the fitness peak the predator previously occupied
has vanished.  The predator might find itself down in a valley instead
of up on a peak.  Then it may start climbing the fitness landscape
again, for instance by developing sharper teeth, until it reaches a
new local peak. This may in turn affect the survivability of other
species, and so on.

Kauffman and Johnsen represented the interdependence of species in
terms of their model.  Mutating one of the N genes in a species
affects K genes within the species and also affects the fitnesses of C
other species.  This is called the NKC model.  Now, starting from a
random configuration, all the species start climbing their landscapes,
and at the same time start modifying each other's landscapes.  Their
idea was that this would eventually set the system up in a poised
state where evolution would happen intermittently in
bursts. However, this failed to occur. 

Either of two scenarios would emerge.  i) If the
number of interactions, C, is small, the ecology evolves to a frozen
state where all  species rest on a local peak of their
respective landscapes, and no further evolution takes place.  A random,
external environment is introduced by randomly flipping genes. This
initiates avalanches where a number of species evolve.  However the
avalanches are small, and the ecology soon comes to rest in a new frozen
state.  ii) If
the the number C is large, the ecology goes to a highly active
chaotic state, where
the species  perpetually evolve without ever reaching a peak.   In this
case, the coevolutionary avalanches never stop.  Only
if the parameter C is carefully tuned does the ecology evolve to a
critical state. 

Kauffman and Johnsen argued that the
ecology as a whole is most fit at the critical point. ``The critical
state is a nice place to be,'' Kauffman claims.  However, it can be
proven that the NKC models do not self-organize to the critical
point (Flyvbjerg and Lautrup 1992; Bak, Flyvbjerg, and Lautrup 1992).
Divine intervention is needed.  Apart from the question as to
what type of dynamics may lead to a critical state, the idea of a
poised state operating between a frozen and a disordered, chaotic
state makes an appealing picture for the emergence of complex
phenomena. A frozen state cannot evolve. A chaotic state cannot
remember the past. This leaves  the critical state as the only
alternative.

\section{A Simple  Model for Evolution}

Bak and Sneppen (Bak and Sneppen 1993; Sneppen et al 1995)
introduced a simple model to describe the main features of
a global interactive ecology. In one version of the model, $L$ species
are situated on a one dimensional line, or circle.  Each species
interacts with its two neighbors, to the left and to the right.  The
system can be thought of as a long food chain.  Instead of specifying
the fitness in terms of a specific algorithm, the fitness of the
species at position $i$ is simply given as a random number, $f_i$, say
between 0 and 1. The fitness is not specified explicitly in terms of
an underlying genetic code, but is chosen as a random function of
these variable.  Probably not much is lost since we do not know the
actual fitness landscapes anyway.

At each time step the least fit species is selected for mutation or
extinction.  This is done by finding the smallest random number in the
system. By inspection of the fitness landscape in figure 3, it is
clear that species located on low fitness peaks have a smaller
distance to find better peaks than species with higher fitness.  The
barriers to find better peaks can be thought of as the number of
coordinated mutations needed.  So the time it takes to reach a higher
peak increases exponentially with the size of the barriers and
can become astronomically large if the genetic mutation rate is
low.  This
justifies the selection of the least fit species as the next in line
for mutation.  

The mutation of a species is represented by replacing
its random number with a new random number.  One might argue that the
new random number should be higher than the old one, but this does not
change the behavior, so for mathematical simplicity we replace the old
random number with a completely new random number between 0 and 1.
One might think of this elementary event either as an extinction of a
species occupying a certain ecological niche followed by the
replacement with another species, or as a pseudo-extinction where a
species mutates.  As far as our mathematical modeling is concerned,
this doesn't make any difference.  The mutation of the species at site
$i$ results in a change in the physical properties of that species, so
that it affects the fitnesses of its two neighboring species. For
simplicity, this is modelled by choosing a new, randomly selected,
fitness for the neighbors also. One might argue their fitness should
only be affected slightly, say less than $1/10$, or that their fitness should
generally be worsened. Again, the details of the model do not affect
the resulting outcome, so we choose a completely new random number.

To summarize: {\it At each time step in the simulation,
the smallest fitness, and the 
fitness of the two neighbors are each replaced with new random fitnesses. 
This step is repeated again and again. That's all!}

What could be simpler than replacing some random numbers with some
other random numbers? Despite the simplicity, the outcome of the
process is nontrivial.  One might suspect that the model
would be so simple that it would easily yield to  mathematical
analysis, but the complexity of its behavior sharply contrasts with its simple
definition. We shall see that modified versions of the model are more tractable.

In particular, a multi-trait evolution model (Boettcher and Paczuski
1996), which behaves similarly to the Bak-Sneppen model, can be
completely solved.  Instead of each site $i$ having a single fitness
$f_i$ it has many fitnesses associated with its $M$ different traits
that evolve independently.  The introduction of many internal traits
is consistent with paleontological observations indicating that
evolution within a species is ``directed''; morphological change over
time is concentrated in a few traits, while most others traits of the
species are static (Kaufmann 1993).  The multi-trait model includes
the Bak-Sneppen model when $M=1$ and is solvable when
$M\rightarrow\infty$.

\subsection{The Self Organized Critical State}
 
Figure 4 shows a snapshot of the fitnesses after billions of 
updates for a Bak-Sneppen ecology with 300 species. Most of the fitnesses are 
above a critical value, or gap, $f_{c} = 0.67002$ (Paczuski, Maslov, and Bak
1996).   Note however a localized 
area in the species space where the fitnesses are below this gap. The 
next species to mutate is the one with the lowest fitness, \#110. The 
fitness of this species and its two neighbors are updated. It is very 
likely that the next species to mutate is nearby.
Subsequent evolution takes place within this localized burst. After a while, 
there will be no more species below the gap, and a new burst, or 
avalanche, will start somewhere else in the system.

During the avalanche, the species are mutating again and again, 
until eventually a self-suspended, stable network of species has been 
reestablished. The size, $s$, of the burst can be measured as the total 
number of mutations separating instances with no species in the gap. 
Figure 5 shows the distribution of burst sizes. There are avalanches 
of all sizes, with a power law distribution

\begin{equation}
P(s) \sim s^{-\tau} \mbox{  where }  \tau \simeq 1.07 \quad .
\end{equation}

The power law for large sizes
shows that the ecology has self-organized to a critical
state. The large avalanches represent mass extinction events, like the
Cambrian explosion (Gould 1989). During the large avalanches,
nature tries one combination after another until a relatively stable
configuration is reached, and a period of stasis in this area of the
global ecology begins.

As a consequence of the interaction between species, even species that
possess well-adapted abilities, with high barriers, can be undermined
in their existence by weak species with which they interact.  For
instance, a species with fitness above the critical gap never mutates
on its own.  However, eventually it may be hit by ``bad luck'' because
a mutation among its neighbors destroys its pleasant and stable life.
A species can go extinct through no fault of its own and for no
apparent external ``reason'' such as a meteor.  Nature is not fair! A
high fitness is only useful as long as the environment defined by the
other species in the ecology remains intact.

 Figure 6 shows the accumulated number of mutations of an arbitrary
single species in the model. Note the relatively long periods of
stasis, interrupted by bursts of activity. One might imagine that the
amount of morphological change of a species is proportional to the
the total number of mutations, so the curve shows punctuated equilibrium
behavior. The large jumps represent periods where very many subsequent
mutations take place at a rapid pace, because an ecological
co-evolutionary avalanche happens to visit the species. Thus, the big
jumps between ``useful'' or highly fit states are effectuated by 
cumulative small jumps through intermediate states which could exist
only in the temporary environment of a burst. The curve is a Cantor
set, or Devil's staircase, invented by the mathematician Georg Cantor
in the last century. The length of the period of stasis is the ``first
return time''  of the activity for a given species. That quantity
also has a power law distribution
(Maslov, Paczuski, and Bak 1995). 
 In evolution, this time can be
thought of as the lifetime of a species before it undergoes a
(pseudo)extinction.   In fact, the distribution of lifetimes of fossil
genera (Sepkoski 1993) appears to follow a power law with a
characteristic exponent $\simeq 2$ (Sneppen et al. 1995).

As a consequence of the power law distribution of burst sizes, 
most of the evolutionary activity occurs within the few large avalanches
rather than the many smaller ``ordinary'' events. Self-organized criticality
can thus be thought of as a theoretical justification for catastrophism.

\subsection{Comparison with the fossil record}
The time unit in the computer simulations is a single update step.  Of
course, this does not correspond to time in real biology.  Based on
the rugged fitness landscape picture, the time-scale for traversing a
barrier of size $b$ is exponential in the barrier height, which is
roughly proportional to the fitness, so $t_i \sim \exp(f_i/T)$. 
Here $T$ is an effective temperature which represents an average mutation
rate in the genetic code.  In real
biology, there is no search committee locating the least fit species,
but mutation takes place everywhere at the same time. In the limit where
the effective temperature $T$
 approaches zero, the minimal fitness is always selected next
for mutation.  Punctuated equilibrium behavior can exist only
where the mutation rate is slow; otherwise there would not be
long periods of stasis.  A system with a high mutation rate will
not have sufficient memory to develop complex behavior, since any new
development will be erased in a relatively short time span. 

Sneppen et al. (1995) performed a simulation where at each time step a 
mutation takes place everywhere with probability $p=exp(-f_i/T)$. 
Figure 7 shows the resulting space-time plot of the activity, with 
$T=0.01$. Note the temporal separation of the avalanches, which show 
up as connected black areas. 
The information in this diagram can be presented differently. In 
figure 8, the time scale has been coarse grained in a simulation over 
8000 steps into 60 equal time intervals. The total amount 
of events in each time step is plotted as a function of time.
 Note the similarity 
with Sepkoski's plot, figure 1. During each time period, there are generally 
many avalanches. One can show that the resulting distribution for 
the total number of events in each period approaches a Pareto-Levy 
distribution, which has power law tails for large events.
The information in Raup's 
histogram of Sepkoski's data is too sparse to test whether or not it 
represents a Pareto-Levy distribution.

In the Bak-Sneppen model, the number of species is conserved.
It does not allow for speciation where one species branches 
into two species, but might be justified as a consequence of competition
for resources. Vandewalle
and Ausloos (1995) have constructed a model where phylogenetic trees are
formed by speciation, starting from a single species. This model also
evolves to the critical state, with extinction events of all sizes.

Of course, it would be interesting if one could perform controlled
 experiments on evolution. The second best is to construct  artificial
evolution in the computer. Ray (1992) and Adami (1995) have done just that.
They created a world of replicating and mutating program segments. Adami
found that evolution in this artificial world exhibits punctuated
equilibria, with a lifetime distribution of organisms following a power
law, indicating that the system self-organizes into a critical state.

\subsection{External Effects}
To what extent is evolution described by such simple models
affected by external events, such as temperature variations, volcanic 
eruptions, neutrino bursts, or meteors? Schmultzi and Schuster (1995) studied a model
where extinction takes place when 
the fitness of a species falls below a random number drawn from an independent 
distribution. The random number represents the effect from external 
sources. Self-organized criticality and punctuated equilibria were 
also found.  Newman and Roberts (1995) 
took a similar approach. The species were assigned not only barriers 
for spontaneous mutation, but independent random fitnesses. At 
each point in time, species with fitnesses less than a randomly 
varying external perturbation would go extinct, in addition to species 
mutating spontaneously. The external perturbations initiate 
avalanches. They found a power law distribution of co-evolutionary 
avalanches, with an exponent $\tau \simeq 2$.

\section{Theory}

Theoretical developments to mathematically describe the behavior of
these abstract computer models have taken two routes.  The first is a
phenomenological approach that we have undertaken in collaboration
with Sergei Maslov (Paczuski, Maslov, and Bak 1994, 1996).  We have
made a unified theory of avalanche dynamics which treats not only
evolution models but also other complex dynamical systems with
intermittency such as invasion percolation, interface depinning, and
flux creep.  Complex behavior such as the formation of fractal
structures, $1/f$ noise, diffusion with anomalous
Hurst exponents, Levy flights, and punctuated equilibria can all be
related to the same underlying avalanche dynamics.  This dynamics can
be represented as a fractal in $d$ spatial plus one temporal
dimension.  In particular, the slow approach to the critical
attractor, i.e.  the process of self-organization, is governed by a
``gap'' equation for the divergence of avalanche sizes as the gap in
the fitness distribution approaches the critical value, starting from
the flat gapless initial distribution.  Figure 9 shows the minimum
value of the fitness vs. the number of update steps. The envelope
function of this curve is the gap.  Avalanches of activity below the
gap separate the instances where the gap grows. Clearly, the
avalanches become bigger and bigger as time goes on.

We have developed a scaling theory that
relates many of the
critical exponents describing various physical properties 
to two basic exponents characterizing
 the fractal attractor. The phenomenological theory does not provide
information about the values of those exponents.
  
The second approach has been aimed at obtaining exact results for
specific models.  For the multitrait model (Boettcher and Paczuski
1996), an explicit equation of motion for the avalanche dynamics can
be derived from the microscopic rules.  Exact solutions of this
equation, in different limits, proves the existence of simple power
laws with critical exponents that verify the general scaling relations
mentioned above.  Punctuated equilibrium is described by a Devil's
staircase with a characteristic exponent 
$\tau_{\mbox{\scriptsize FIRST}}=2 - d/4$ where $d$ is the spatial dimension.  Actually, for
the multi-trait evolution model, the distribution of avalanche sizes
is known exactly when $M\rightarrow\infty$.  It is
\begin{equation}
P(s)= {\Gamma\left(s+{1\over 2}\right)\over
\Gamma\left(1\over 2\right)\Gamma\left(s+2\right)}
\sim s^{-3/2} \mbox{ for } s\gg1  \quad .
\end{equation}

This distribution of sizes is the same as for the random neighbor models
in which each species interacts with 2 randomly chosen other species in the 
ecology rather than with near neighbors on a regular grid (Flyvbjerg et al.
1993; deBoer et al. 1995). The power law has a characteristic exponent
$\tau=3/2$ rather than $\tau=1.07$ for the Bak-Sneppen chain.

In the multitrait model, avalanches propagate via a ``Schr\"odinger''
equation in imaginary time with a nonlocal potential in time.  This
nonlocal potential gives rise to a non-Gaussian (fat) tail for the
subdiffusive spreading of activity.   For the chain,
the probability for the activity
to spread beyond a species distance $r$ in time $s$ decays as
$\sqrt{24\over\pi}s^{-3/2}x \, \exp{[-{3\over 4}x]}$ for $x=({r^4\over
s})^{1/3} \gg 1$ (Paczuski and Boettcher 1996).  This anomalous
relaxation comes from a hierarchy of time scales, or memory effect,
that is generated by the avalanches.
In
addition, a number of other correlation functions characterizing the
punctuated equilibrium dynamics have been determined exactly.  For
instance, the probability for a critical avalanche to affect a species
at distance $r$ is exactly $12/((r+3)(r+4))$ in one dimension.

\section{Acknowledgments}

This work is supported by the Division of Materials Science, U. S. 
Department of Energy under contract \# DE-AC02-76CH00016. We are 
grateful for discussions and collaborations leading to the results 
summarized here with K. Sneppen, H. Flyvbjerg, and S. Maslov.
We thank Mike Fuller for biologically relevant comments on our paper.

\begin{figure}
\vspace{5 cm}
\caption[]{Temporal pattern of extinctions recorded over the last 600 
million years, as given by J. Sepkoski (1993). The ordinate shows 
an estimate of the percentage of species that went extinct within 
intervals of 5 million years}
\end{figure}

\begin{figure}
\vspace{8 cm}
\caption[]{Histogram of extinction events as shown by Raup (1986).
The histogram is based on the recorded time of extinction of 2316
marine animal families}
\end{figure}

\begin{figure}
\vspace{5 cm}
\caption[]{Fitness landscape. Note that the species with low fitnesses
have smaller barriers to overcome in order to improve their fitness
than species with high fitnesses}
\end{figure}

\begin{figure}
\vspace{5 cm}
\caption[]{Snapshot of fitnesses $f$ for 300 species during an avalanche in the
evolution model. Most of the $f$ values are above the critical value. The
cluster of active species with $f<f_c$ participate in the avalanche and
undergo frequent changes (Paczuski et al. 1996)}
\end{figure}

\begin{figure}
\vspace{5 cm}
\caption[]{Distribution of the size of avalanches in the critical state for
the one dimensional evolution model. The straight line on the log-log
plot indicates a power law with an exponent $\tau \simeq 1.07$
 (Paczuski et al. 1996)}
\end{figure}

\begin{figure}
\vspace{5 cm}
\caption[]{ Accumulated number of mutations for a single species, or a
single ecological niche in the stationary state. The curve exhibit
punctuated equilibrium behavior, with periods of stasis interrupted
by intermittent bursts. The curve is a Cantor set, or a Devil's
staircase}
\end{figure}

\begin{figure}
\vspace{5 cm}
\caption[]{Space time plot of the activity. The horizontal axis is the
species axis. The time at which a species mutates is shown as a black dot. The
avalanches appear as connected black areas. Calculation was done for
a value of the mutation parameter $T=0.01$ (Sneppen et al. 1995)}
\end{figure}

\begin{figure}
\vspace{5 cm}
\caption[]{Temporal pattern of evolution, with
$T=0.01$. Note the similarity with Sepkoski's plot, figure 1
 (Sneppen et al. 1995)}
\end{figure}

\begin{figure}
\vspace{5 cm}
\caption[]{The self-organization process for a small system. 
$f_{\mbox {\scriptsize min}}$
vs time is shown (crosses). The full curve shows the gap, which is the
envelope function of $f_{\mbox {\scriptsize min}}$. On average, the
avalanche size grows as the critical point is approached, and
eventually diverges, as the gap approaches the critical value
0.6700. (Paczuski et al. 1996)}
\end{figure}


\begin{thebibliography}

\bibitem{}{adami}{}
Adami, C. (1995): Self-organized criticality in living systems.
Phys. Lett. A {\bf 203}, 29--32

\bibitem{}{alvarez}{Alvarez, Alvarez, and Michel (1980)}
Alvarez, L. W., Alvarez, F. A., Michel, H. W. (1980):
Extraterrestrial cause for the Cretaceous--Tertiary extinction
Science {\bf 208},
1095--1108

\bibitem{}{bfl}{}Bak, P., Flyvbjerg, H., Lautrup, B. (1992): Coevolution in a
rugged fitness landscape. Phys. Rev. A
 {\bf 46}, 6724--6730

\bibitem{}{bp}{Bak and Paczuski (1995)}
Bak, P., Paczuski M. (1995):
Complexity, contingency, and criticality.
Proc. Natl Acad. Sci. USA {\bf 92} 6689-6696

\bibitem{}{bs}{Bak and Sneppen (1993)}
Bak, P., Sneppen, K. (1993):
Punctuated equilibrium and criticality in a simple model of evolution.
Phys. Rev. Lett. {\bf 71}, 4083--4086

\bibitem{}{btw1}{Bak, Tang, and Wiesenfeld (1987)}
Bak, P., Tang, C., Wiesenfeld K. (1987):
Self-organized criticality. An explanation of 1/f noise.
Phys. Rev. Lett. {\bf 59}, 381--384

\bibitem{}{btw2}{Bak, Tang, and Wiesenfeld (1988)}
Bak, P., Tang, C., Wiesenfeld K. (1988):
Self-organized Criticality.
Phys. Rev. A {\bf 38}, 364--374

\bibitem{}{bp}{}
Boettcher S., Paczuski M. (1996): Exact results for spatiotemporal
correlations in a self-organized critical model of punctuated equilibria.
 Phys. Rev. Lett. {\bf 76},348-351 

\bibitem{}{darwin}{Darwin (1910)}
Darwin, C. (1910): {\it The Origin of Species} (Murray, London), 6th
edition.

\bibitem{}{derrida}{deBoer {\it et al} (1994)}
deBoer, J., Derrida, B., Flyvbjerg, H., Jackson, A. D., Wettig, T. (1994):
Simple model of self-organized biological evolution.
Phys. Rev. Lett. {\bf 73}, 906--909

\bibitem{}{fisher}
{}Fisher, R. A. (1932): {\it The Genetical Theory of Natural Selection}

\bibitem{}{fl}{}Flyvbjerg. H., Bak P., Jensen, M. H. Sneppen, K. (1995):
A self-organized Critical Model for Evolution,
in {\it Modelling the Dynamics of Biological Systems}.
E. Mosekilde and O. G. Mouritsen (Eds.)
(Springer Berlin, Heidelberg, New York)pp 269--288

\bibitem{}{fl}{}Flyvbjerg, H., Lautrup, B. (1992): Evolution in a
rugged fitness landscape. Phys. Rev. A
 {\bf 46}, 6714--6723

\bibitem{}{fsb}{Flyvbjerg, Sneppen, and Bak (1993)}
Flyvbjerg, H., Sneppen K., Bak P. (1993):
Mean field theory for a simple model of evolution.
Phys. Rev. Lett. {\bf 71}, 4087--4090

\bibitem{}{gould}{Gould (1989)}
Gould, S. J. (1989): {\it Wonderful Life} (Norton, New York)

\bibitem{}{gould-eldredge}{Gould and Eldredge (1977)}
Gould, S. J., Eldredge N. (1977):
Punctuated equilibria:  The tempo and mode of evolution reconsidered.
Paleobiology {\bf 3}, 114--151

\bibitem{}{gutenber}{Gutenberg and Richter (1956)}
Gutenberg B., Richter C. F. (1956): Ann. Geofis. {\bf 9}, 1

\bibitem{}{Kauffman}{(Kauffman 1993)}
Kauffman, S. A. (1993): {\it The Origins of Order: Self-Organization and
Selection in Evolution} (Oxford University Press, Oxford)

\bibitem{}{kj}{(Kauffman and Johnsen 1991)}
Kauffman, S. A., Johnsen, S. J. (1991):
Coevolution to the edge of chaos:   Coupled fitness landscapes,
poised states, and coevolutionary avalanches.
J. Theo. Biology {\bf 149}, 467--505

\bibitem{}{mpb94}{Maslov, Paczuski, and Bak (1994)}
Maslov, S., Paczuski, M., Bak, P. (1994):
Avalanches and $1/f$ noise in evolution and growth models.
Phys. Rev. Lett. {\bf 73},
2162--2165

\bibitem{}{smith}{ Maynard Smith (1993)}
Maynard Smith, J. (1993): {\it The Theory of Evolution}
(Cambridge University Press, Cambridge) 

\bibitem{}{newman2}{Newman and Roberts (1995)}
Newman, M. E. J., Roberts, B. W. (1995):
Mass extinctions:  Evolution and the effects of external influences on
unfit species.
 Proc. Roy. Soc. London B {\bf 260},
31

\bibitem{}{newman1}{(Newman, Turcotte, and Gabrielov
1995)}
Newman, W. L., Turcotte D. L., Gabrielov A. M. (1995): Log-periodic
behavior of a hierarchical failure model. Phys. Rev. E {\bf 52},
4827

\bibitem{}{pmb1}{Paczuski, Maslov, and Bak (1994)}
Paczuski, M., Maslov, S., Bak, P. (1994):
Field theory for a model of self-organized criticality.
Europhys. Lett. {\bf 27}, 97--100; Erratum
{\bf 28},
 295--296

\bibitem{}{pmb2}{Paczuski, Maslov, and Bak (1996)}
Paczuski, M., Maslov, S., Bak, P. (1996):
Avalanche dynamics for evolution, growth, and depinning models.
Phys. Rev. E {\bf 53}, 414

\bibitem{}{pacb}{Paczuski and Boettcher (1996)}
Paczuski, M., Boettcher, S. (1996): in preparation

\bibitem{}{raup1}{Raup (1982)}
Raup,\, D.M. (1982):{\it Extinction. Bad Genes or Bad Luck?} (Oxford 
University Press, Oxford)

\bibitem{}{raup2}{Raup (1986)}
Raup, D. M. (1986): Biological extinction in history.
 Science {\bf 231}, 1528--1533 

\bibitem{}{ray}{}
Ray, T. (1992): in {\it Artificial Life II} ed. C. G. Langton
(Addison-Wesley, Reading Massachussetts)

\bibitem{}{sc}{}
Schmultzi, K., Schuster, H. G. (1995): Introducing a real time scale into
the Bak-Sneppen model. Phys. Rev. E {\bf 52}, 5273

\bibitem{}{sepkoski}{Sepkoski (1993)}
Sepkoski, J. J. Jr. (1993): 
Ten years in the library: New data confirm paleontological patterns
Paleobiology {\bf 19}, 43--51

\bibitem{}{sbfj}{Sneppen, Bak, Flyvbjerg, and Jensen (1995)}
Sneppen. K. , Bak P., Flyvbjerg H., Jensen, M. H. (1995): 
Evolution as a self-organized critical phenomena.
Proc. Natl. 
Acad. Sci. USA {\bf 92}, 5209--5213

\bibitem{}{va}{}
Vandewalle, N., Ausloos, M. (1995): Self-organized criticality in
phylogenetic-like tree growth. J. de Phys. (Paris) {\bf 5}, 1011--1025

\bibitem{}{wright}{Wright (1982)}
Wright, S. (1982): Character change, speciation,
and the higher taxa.  Evolution {\bf 36}, 427--443

\end{thebibliography}
\end{document}